# Electronic phase separation in the itinerant metamagnetic transition of $Sr_4Ru_3O_{10}$

Z.Q. Mao*, M. Zhou, and J. Hooper
Department of Physics, Tulane University, New Orleans, LA 70118

V. Golub and C. J. O'Connor
Advanced Materials Research Institute, University of New Orleans,
New Orleans, LA 70148

Abstract

Triple-layered ruthenate $Sr_4Ru_3O_{10}$ shows a first-order itinerant metamagnetic transition for in-plane magnetic fields. Our experiments revealed rather surprising behavior in the low-temperature transport properties near this transition. The in-plane magnetoresistivity $\rho_{ab}(H)$ exhibits ultrasharp steps as the magnetic field sweeps down through the transition. Temperature sweeps of $\rho_{ab}$ for fields within the transition regime show non-metallic behavior in the up-sweep cycle of magnetic field, but show a significant drop in the down-sweep cycle. These observations indicate that the transition occurs via a new electronic phase separation process; a lowly polarized state is mixed with a ferromagnetic state within the transition regime.

Electronic phase separation in strongly correlated electron systems has been an important research topic in the past decade since it is associated with colossal magnetoresistance (CMR) in manganites and high-temperature superconductivity (HTSC) in cuprates. The phase separation takes different forms depending on the specific characteristics of the system. CMR manganites phase-separate into a mixture of antiferromagnetic insulating regions and ferromagnetic (FM) metallic domains [1-3], while the phase separation in HTSC cuprates results in the formation of spin and charge stripes [4]. Here we report a new type of phase separation phenomenon observed in the ruthenate $Sr_4Ru_3O_{10}$. This material shows an itinerant metamagnetic transition at about 2 T for magnetic fields applied along the in-plane direction. Our experiments demonstrate that this transition occurs through a mixed lowly-polarized / FM phase process accompanied by quantum critical fluctuations. This inhomogeneous electronic state results in ultrasharp magnetoresistivity steps in the down-sweep cycle of magnetic field, which is very unusual for bulk transport properties.

Perovskite strontium ruthenates of the Ruddlesden-Popper (RP) series $Sr_{n+1}Ru_nO_{3n+1}$ show fascinating physics, including spin-triplet superconductivity in $Sr_2RuO_4$ ($n = 1$) [5-7], metamagnetic quantum criticality in $Sr_3Ru_2O_7$ ($n = 2$) [8-11], and itinerant ferromagnetism in $SrRuO_3$ ($n = \infty$) which shows evidence for a possible magnetic monopole in the crystal momentum space [12]. Considerable investigation has been devoted to these materials in recent years. $Sr_4Ru_3O_{10}$ is the triple-layered member of the RP series with $n = 3$. Although this material has not been as widely studied as the other three members, earlier work by Cao *et al* [13] revealed that its ground state has intriguing characteristics: it is poised between an itinerant metamagnetic state and an itinerant ferromagnetic state. When the field is applied along the *c*-axis typical ferromagnetic behavior occurs, while for field applied along the in-plane direction a first-order metamagnetic transition accompanied by critical fluctuations is observed. The critical field of this transition is much less than that of $Sr_3Ru_2O_7$ [8]. A recent theoretical work [14] interprets this itinerant metamagnetism within a mean-field theory by considering the proximity of the Fermi level to a van Hove singularity. We have performed systematic investigations on the in-plane electronic transport properties near the metamagnetic transition of $Sr_4Ru_3O_{10}$ for fields parallel to the *ab*-plane using high-quality single crystals.

Our crystals were prepared by a floating-zone technique; crystal growth conditions were reported elsewhere [15]. Crystals selected for the measurements were well characterized by x-ray diffraction and found to be pure $Sr_4Ru_3O_{10}$. The transport data presented in this paper were obtained on a crystal with a residual resistivity $\rho_0 = 6.2$ $\mu\Omega$.cm. Our transport measurements were carried out in a $^3$He cryostat (with a base temperature of 0.3 K) using a standard four-probe technique. Magnetization measurements were made with a SQUID magnetometer.

The inset (a) of Fig. 1 shows the magnetization as a function of magnetic field at 2 K for a $Sr_4Ru_3O_{10}$ crystal. Similar to the results reported previously [13], we observed FM behavior for *H//c* and a metamagnetic transition with significant hysteresis for *H//ab*. Near the metamagnetic transition, the in-plane resistivity $\rho_{ab}$ exhibits a sharp change, as shown in the inset (b) of Fig. 1 where $B^+_{c1}$ (~1.75 T), $B^+_{c2}$ (~2.50 T), $B^-_{c1}$ (1.20 T), and $B^-_{c2}$ (~2.00 T) are defined as lower and upper critical fields of the transition for the upwards ($^+$) and downwards ($^-$) sweep cycles of magnetic field. $\rho_{ab}$ displays discrete jumps as the field sweeps down through the transition. These jumps were found to be extraordinarily sharp, as shown in the main panel of Fig. 1 where the data was taken with a step of 1 G in the 1.6-1.2 T field range of the down-sweep cycle. These ultrasharp steps were reproduced on all samples we measured. They are rather surprising and have not been observed in any other itinerant electron system that we know of.

These magnetoresistivity steps are sensitive to the magnetic history. Shown in Fig. 2a are the data taken in different field–sweep cycles using zero-field cooling (ZFC). For each cycle, the field first sweeps from zero up to a certain value (defined as a terminal field $B_T$), then sweeps backwards toward zero. We note that the steps occurring in the downward sweep depend on $B_T$. When $B_T$ is less than the lower critical field $B^+_{c1}$, 1.75 T, no steps occur in $\rho_{ab}$ and only a small hysteresis is observed. When $B_T$ is past $B^+_{c1}$, the steps appear and grow in number with increasing $B_T$.

Figure 2b displays the data measured with various field cooling (FC) histories. Each curve was obtained by sweeping the field down from the given field for FC (defined as $B_{FC}$). We observed that under FC, steps develop even when $B_{FC} < B^+_{c1}$, in sharp contrast with the situation of ZFC. For FC at 1.70 T, we performed three independent measurements using the same cooling history and observed different behavior in the steps. The positions of the steps were shifted slightly between different measurements. This behavior (which was also observed for FC with $B_{FC} > B^+_{c1}$ and ZFC with $B_T > B^+_{c1}$) reflects the non-equilibrium character of the steps. Further increasing $B_{FC}$ results in a greater number of steps. These steps also depend on temperature, disappearing above 1 K.

What is the origin of these magnetoresistivity steps? Our experiments reveal that they are associated with a mixed phase process occurring within the metamagnetic transition. As seen in inset (a) of Fig. 1, the metamagnetic transition for $H//ab$ corresponds to a super-linear increase in magnetization at about 2 T. The magnetization increases almost linearly with field in the lower field region below the transition, and this increase is much slower than the situation for $H//c$. Although the mechanism for this significant magnetic anisotropy is unclear, we define the system in this low field range ($B<B^+_{c1}$) for $H//ab$ as a lowly polarized (LP) state following the normal definition of itinerant metamagnetism. For high fields above the transition ($B>B^+_{c2}$), the saturation of both magnetization and magnetoresistivity clearly indicates that the system is nearly fully polarized. We define the system as a forced ferromagnetic (FFM) state for this field regime. Near or within the transition, our data demonstrate that the system exists as a mixture of LP and FFM phases.

Since the FFM phase has lower resistivity than the LP phase (see the inset (B) of Fig. 1), the LP/FFM mixed phase process manifests itself through very unique characteristics as stated below. (1) Under ZFC, the downward field-sweep of $\rho_{ab}$ shows very unusual hysteretic behavior for $B_T$ slightly above $B^+_{c1}$ (e.g., see the data taken with $B_T = 1.90$ in Fig. 2a). $\rho_{ab}$ first decreases linearly, similar to the LP state for $B_T < B^+_{c1}$; it then increases prominently after a broad minimum and subsequently shows discrete jumps. This behavior can best be understood as resulting from an inhomogeneous phase with FFM domains embedded within a LP matrix. The initial linear decrease indicates that the dominant contribution to transport is the LP phase, while the prominent increase below the minimum, as well as the significant hysteresis, reflects contributions from the FFM phase which must transit to the LP phase at sufficiently low fields. The presence of steps indicates that the LP phase forms continuous domain walls in the down-sweep cycle (see below). As $B_T$ increases, we can easily imagine that the volume ratio of FFM to LP should increase. FFM domains should form a percolative network and dominate transport properties when $B_T$ approaches $B^+_{c2}$. This conjecture is fully consistent with our observation: the slope of the initial linear part in the downward sweep of $\rho_{ab}$ (which reflects the contribution of the LP phase to transport) decreases gradually with increasing $B_T$ and becomes zero as $B_T$ approaches $B^+_{c2}$ (e.g. see the data taken with $B_T = 2.07$ T).

(2) Under FC, as seen in Fig. 2b, although $B_{FC} < B^+_{c1}$, the downward sweep of $\rho_{ab}$ behaves similarly as that for ZFC with $B^+_{c1} < B_T < B^+_{c2}$, i.e., it shows a minimum and steps (e.g. see the data taken with $B_{FC} = 1.55$, 1.65 and 1.70 T). This observation further supports the mixed phase picture near the transition as proposed above. The FC process favors FFM domains and would increase the volume ratio of FFM to LP phase. $\rho_{ab}$ therefore exhibits the properties of the mixed phase even for $B_{FC} < B^+_{c1}$.

Furthermore, evidence for phase separation near the transition was also observed in the temperature sweep of $\rho_{ab}$. Figure 3a shows $\rho_{ab}(T)$ measured at various fields for the up-sweep cycle of magnetic field under ZFC. $\rho_{ab}$ displays a remarkable non-metallic temperature dependence below 5 K for $B^+_{c1} < B < B^+_{c2}$, in sharp contrast with the behavior seen outside the transition regime where $\rho_{ab}$ shows $T^2$ dependence for $B < B^+_{c1}$, and $T^{5/3}$ dependence for $B > B^+_{c2}$. This observation is surprising, but fits very well into the above picture of phase separation. As indicated above, when the field is past $B^+_{c1}$ for ZFC, prominent FFM domains should develop. Before these domains are continuously connected to form a percolative network, scattering at domain boundaries would have a substantial influence on transport properties. The experimental facts listed below strongly suggest that this non-metallic temperature dependence originates from this domain-boundary scattering.

First, this non-metallic behavior tends to diminish as the field approaches $B^+_{c2}$, disappearing above $B^+_{c2}$. This observation agrees with the expectation that FFM domains should develop to form a network in this higher field regime and the percolative transport through this network would be dominant. Domain boundary scattering would not be involved in the transport process in this case. This view is further supported by the observation that the temperature dependence of $\rho_{ab}$ is less sensitive to the field for $B > B^+_{c2}$ (for example, see the data taken at 2.5 and 6.0 T, both showing $T^{5/3}$ dependence).

Secondly, as seen in Fig. 3b, the non-metallic behavior can be suppressed by FC. It has been pointed out above that the FC process favors FFM domains and increases the volume fraction of the FFM phase. Thus percolative transport through the FFM phase would occur at lower fields under FC. Our data clearly suggest that the FFM percolative network starts to form at about $B^+_{c1}$ (1.75 T) under FC, in contrast with the situation seen under ZFC for which a percolative network forms as the field approaches $B^+_{c2}$. In addition, we noted that the irreversibility in $\rho_{ab}(T)$ between ZFC and FC actually develops starting at 1.4 T, implying that FFM domains start to develop below $B^+_{c1}$.

We also measured temperature sweeps of $\rho_{ab}$ in the down-sweep cycle of magnetic field under ZFC, as shown in Fig. 3c. Surprisingly, $\rho_{ab}(T)$ shows a striking drop within the same field range (1.2-1.6 T) where the field sweep of $\rho_{ab}$ exhibits ultrasharp steps, in sharp contrast with the non-metallic temperature dependence seen in the transition range of the up-sweep cycle. This disparity can be understood in the following way. When the field sweeps down from above $B^+_{c2}$, the initial state of the system is FFM and LP domains should develop as the field approaches the transition. When the field enters the $B^-_{c1}$ (1.2 T) $< B <$ $B^-_{c2}$ (2.0 T) transition region, the volume fraction of LP should increase prominently. LP domains should be connected to form continuous walls when the field is low enough (< 1.6 T); this can account for the presence of the ultrasharp magnetoresistivity steps (see below for further discussions). Since the volume ratio of the FFM/LP mixed phases is temperature dependent, the observation of a resistivity drop in the temperature sweep during the down-sweep cycle can be attributed to an increase of the volume ratio of the FFM phase caused by decreasing temperature. This volume increase would restore FFM percolative transport. This point of view is supported by the fact that the decrease of temperature at 1.55 or 1.60 T causes $\rho_{ab}$ to decrease to the same value as that for high applied fields, where the transport is dominated by a FFM percolation network.

We have fit our $\rho_{ab}(T)$ data in the 5-12 K range obtained in the down-sweep cycle (including the data not shown in Fig. 3c) to $\rho = \rho_0 + AT^n$. The coefficient $A$ and the exponent $n$ extracted from these fittings are shown in Fig. 3d. The divergence of $A$ at 1.75 T, as well as the sharp change of $n$, indicates that the transport properties are affected by critical fluctuations in the high temperature range where no complete percolation network exists, consistent with earlier work which reported evidence of fluctuations in *c*-axis resistivity [13]. However, for the up-sweep cycle, domain boundary scattering dominates the transport properties for fields within the transition, smearing out the effects of the critical fluctuations.

The schematic shown in Fig. 4 summarizes the mixed phase process discussed above. For the up-sweep cycle with ZFC, the field starts to induce FFM domains below $B^+_{c1}$; the transport properties are dominated by the LP phase until $B$ reaches 1.75 T. When the field is past 1.75 T, but less than 2.5 T, FFM domains grow remarkably and domain-boundary scattering dominates the transport properties at low temperature (see Fig. 4a), resulting in a non-metallic temperature dependence in $\rho_{ab}$. As the field approaches 2.5 T, FFM domains form a percolative network which dominates the transport properties (see Fig. 4b). For the down-sweep cycle, LP domains develop as the field enters the 2.0-1.6 T field range (see Fig. 4c); transport properties are still dominated by the FFM matrix. As the field further sweeps down, below 1.6 T, LP domains with higher resistivity are spontaneously connected and form continuous domain walls at certain threshold fields (see Fig. 4d), which hinders the current flow in the FFM matrix and results in ultra-sharp steps in magnetoresistivity. The disappearance of the steps above 1 K suggests that these LP domain walls are sensitive to thermal excitations.

The dynamic behavior of the steps observed under different cooling histories, shown in Fig. 2, can all be understood within this picture of a mixed phase process. The non-equilibrium character of the steps suggests that these domain walls are not static but fluctuating, as would be expected. With the increase of $B_T$ for ZFC or $B_{FC}$ for FC, the volume fraction of the FFM phase would increase, thus leading to an increase in the number of LP domain walls as the field sweeps down through the transition. Therefore, the number of steps tends to grow with increasing $B_T$ or $B_{FC}$.

In summary, our experiments reveal that the metamagnetic transition of $Sr_4Ru_3O_{10}$ for in-plane fields occurs through a mixed phase process. This result provides key evidence for a recent theoretical model [16] proposed to interpret the observation of a new phase near the metamagnetic quantum phase transition in $Sr_3Ru_2O_7$ [11]. This theory indicates that dipolar magnetostatic forces in an itinerant metamagnet can lead to the formation of unique magnetic domains (called Condon domains). Condon domains differ by the amount of magnetization rather than by direction; domain-wall scattering is expected to be very strong in this situation, which could result in increased resistivity. This expectation is fully consistent with our observations.

Finally, we would like to point out that the electronic phase separation we observed in $Sr_4Ru_3O_{10}$ is intrinsically different from that in CMR manganites, where sharp magnetoresistivity steps were also observed in the up-sweep of magnetic field [17]. The phase separation in manganites occurs between an AFM insulating phase and a FM metallic phase as indicated above, while the phase separation in $Sr_4Ru_3O_{10}$ appears in the form of two mixed metallic phases with different magnetizations.

We would like to thank Drs. W. Bao, B. Binz, Y. Liu, Y. Maeno, M. Norman, P. Schiffer, M. Sigrist, and I. Vekhter for useful discussions, and D. Fobes for technical support. This work was supported by the Louisiana Board of Regents support fund and the George Lurcy fund.

*Electronic address: zmao@tulane.edu.

Figure captions:

Figure 1: Downward field sweep of in-plane resistivity $\rho_{ab}$ for $H//ab$. Inset (a): Magnetization as a function of magnetic field for both $H//c$ and $H//ab$. Inset (b): Field sweeps of $\rho_{ab}$ within the 0-3 T range for $H//ab$. $B^+_{c1}$ (up-sweep) and $B^-_{c1}$ (down-sweep) are defined as the lower critical fields of the transition where $\rho_{ab}$ shows a peak; $B^+_{c2}$ (up-sweep) and $B^-_{c2}$ (down-sweep) are defined as the upper critical fields above which $\rho_{ab}$ tends to saturate.

Figure 2: Field sweeps of $\rho_{ab}$ of $Sr_4Ru_3O_{10}$ under various ZFC (a) and FC (b) histories. $B_T$ is the terminal field for each sweep cycle for ZFC. The data for FC at 1.70, 1.90 and 9.00 T in (b) are shifted for clarity.

Figure 3: (a) $\rho_{ab}(T)$ at $\mu_0 H = 1.60, 1.75, 1.82, 1.89, 1.92, 2.07, 2.50, 6.00$ T in the up-sweep cycle of magnetic field. Each curve was measured by warming up after first cooling the sample down to 0.3 K under zero field and then increasing the field to the target value. (b) $\rho_{ab}(T)$ at various fields for the up-sweep cycle under FC (blue curves); data for ZFC (red curves) are included for comparison. Data are shifted for clarity. (c) $\rho_{ab}(T)$ at $\mu_0 H = 1.20, 1.27, 1.40, 1.50, 1.55, 1.60, 1.70, 1.75, 1.90$ T in the down-sweep cycle. The data were taken by warming up after first cooling the sample down to 0.3 K under zero field, then increasing the field up to 3 T to enter the fully polarized state, and subsequently decreasing the field down to the target value. (d) Parameters extracted from fitting the data in (c) to $\rho = \rho_0 + AT^{\,n}$ (see the text).

Figure 4: Schematics of the LP/FFM mixed phase process within the metamagnetic transition of $Sr_4Ru_3O_{10}$ for both the up-sweep (left column) and the down-sweep (right column) cycles. LP: lowly polarized state; FFM: forced ferromagnetic state. For the up-sweep cycle, FFM domains do not form a percolative network until the field approaches $B^+_{c2}$. For the down-sweep cycle, LP domains form continuous walls when the field is below 1.6 T.

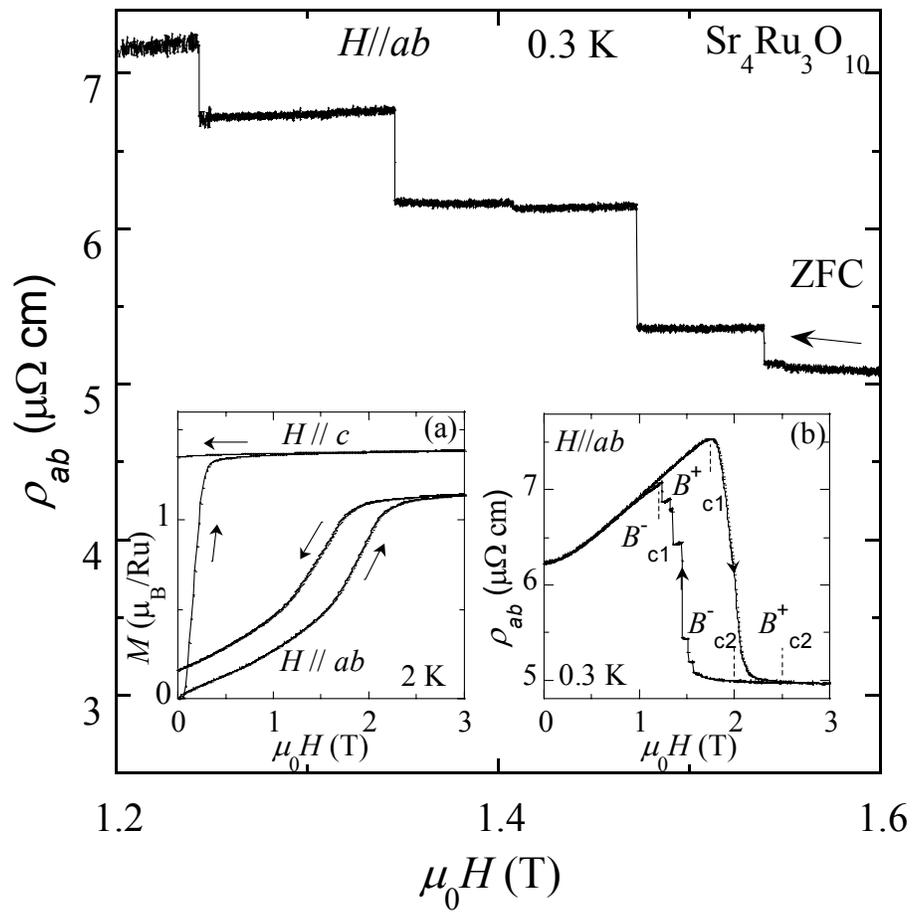

Mao et al., Figure 1.

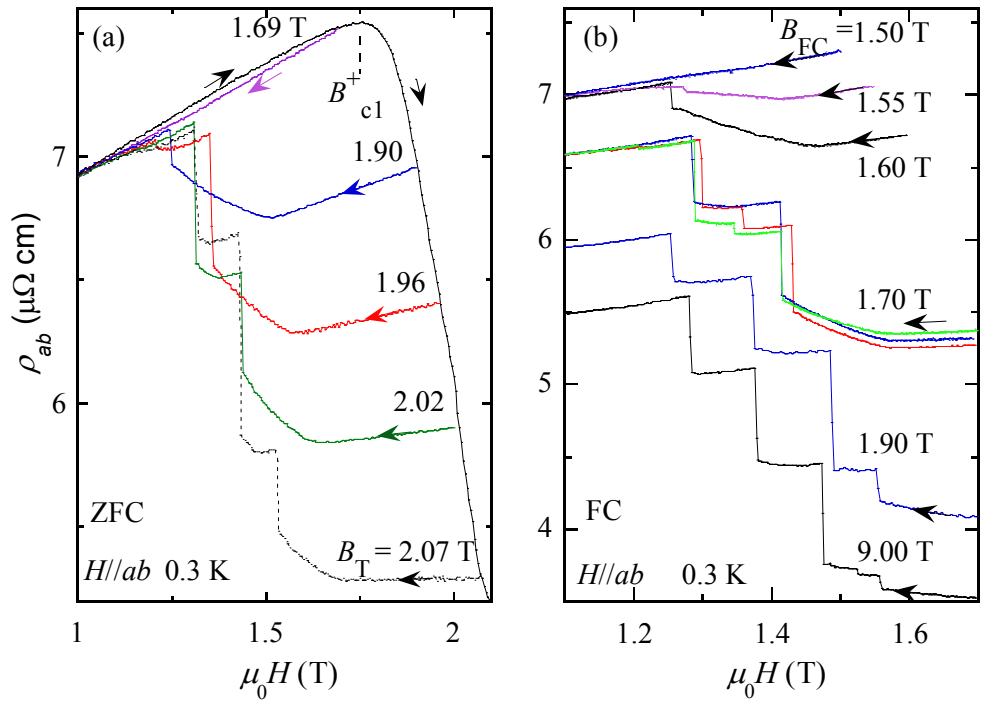

Mao et al., Figure 2

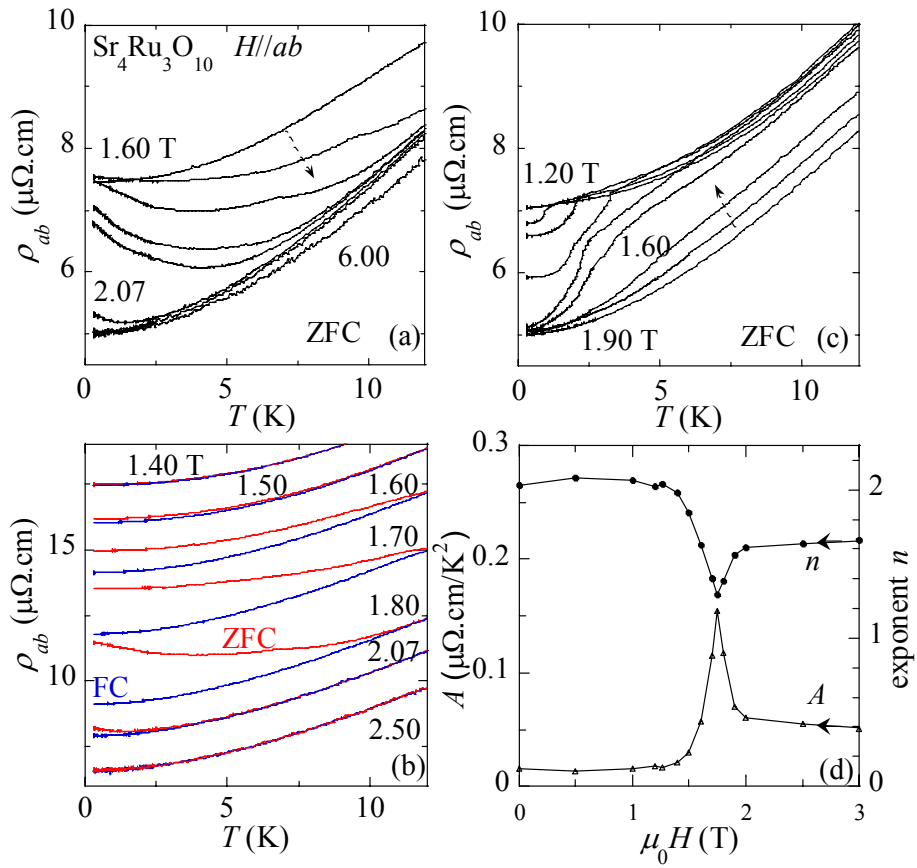

Mao et al., Figure 3

| Up-sweep | Down-sweep |
|---|---|
| 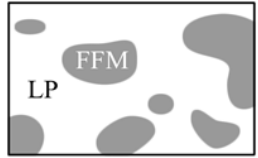 | 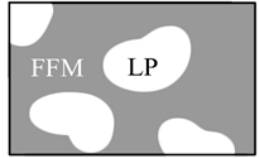 |
| (a) 1.75 T < $B$ < 2.50 T | (c) 1.60 T < $B$ < 2.00 T |
| 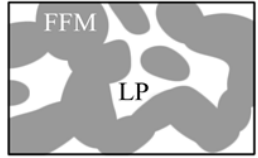 | 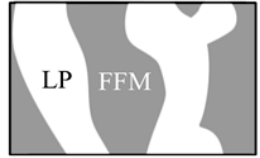 |
| (b) $B$ ~ 2.50 T | (d) 1.20 T < $B$ < 1.60 T |

**Mao et al., Figure 4**